\documentclass{article}
\usepackage[utf8]{inputenc}

\usepackage{subfig}
\usepackage{graphicx}

\usepackage{authblk}
\title{Online Hate: Behavioural Dynamics and Relationship with Misinformation}

\author[1]{Matteo Cinelli}
\author[2,3]{Andra\v{z} Pelicon}
\author[2]{Igor Mozeti\v{c}}
\author[4]{Walter Quattrociocchi}
\author[2]{Petra Kralj Novak}
\author[1,*]{Fabiana Zollo}
\affil[1]{Ca' Foscari University of Venice, Italy}
\affil[2]{Jozef Stefan Institute, Ljubljana, Slovenia}
\affil[3]{Jozef Stefan International Postgraduate School, Ljubljana, Slovenia}
\affil[4]{Sapienza University of Rome, Italy}
\affil[*]{Corresponding author: fabiana.zollo@unive.it}
\date{}

\usepackage{multirow}
\usepackage{graphicx}
\usepackage{appendix}
\usepackage{hyperref}

\usepackage{subfig}

\usepackage{soul}
\usepackage{color}

\newcommand{\acc}{$\mathit{Acc}$}
\newcommand{\fone}{$F_{1}$}
\newcommand{\alfa}{$\mathit{Alpha}$}

\begin{document}

\maketitle

\section*{Abstract}
Online debates are often characterised by extreme polarisation and heated discussions among users. The presence of hate speech online is becoming increasingly problematic, making necessary the development of appropriate countermeasures. In this work, we perform hate speech detection on a corpus of more than one million comments on YouTube videos through a machine learning model fine-tuned on a large set of hand-annotated data. Our analysis shows that there is no evidence of the presence of ``serial haters'', intended as active users posting exclusively hateful comments. 
Moreover, coherently with the echo chamber hypothesis, we find that users skewed towards one of the two categories of video channels (questionable, reliable) are more prone to use inappropriate, violent, or hateful language within their opponents community. Interestingly, users loyal to reliable sources use on average a more toxic language than their counterpart. Finally, we find that the overall toxicity of the discussion increases with its length, measured both in terms of number of comments and time. Our results show that, coherently with Godwin's law, online debates tend to degenerate towards increasingly toxic exchanges of views.

\section{Introduction}
Public debates on social media platforms are often heated and polarised~\cite{adamic2005political, flaxman2016filter, coe2014online}.
Back in the 90s, Mike Godwin coined a theorem, today known as Godwin's law, stating that “As an online discussion grows longer, the probability of a comparison involving Nazis or Hitler approaches to one". More recently, with the advent of social media, an increasing number of people is reporting exposure to online hate speech~\cite{siegel2019online}, leading institutions and online platforms to investigate possible solutions and countermeasures~\cite{gagliardone2015countering}. To prevent and counter the spread of hate speech online, for example, the European Commission agreed with Facebook, Microsoft, Twitter, YouTube, Instagram, Snapchat, Dailymotion, Jeuxvideo.com, and TikTok a “Code of conduct on countering illegal hate speech online”\footnote{\url{https://ec.europa.eu/newsroom/just/document.cfm?doc_id=42985}}. 
In addition to fuelling the toxicity of the online debate, hate speech may have severe offline consequences. Some researchers hypothesised a causal link between online hate and offline violence~\cite{calvert1997hate,chan2016internet,muller2018fanning}. Furthermore, there is empirical evidence that online hate may induce fear of offline repercussions~\cite{awan2015we}. However, the detection and contrast of hate speech is complicated. There are still ambiguities in the very definition of hate speech, with academic and relevant stakeholders providing their own interpretation~\cite{siegel2019online}, including social media companies such as Facebook\footnote{\url{https://www.facebook.com/communitystandards/introduction}}, Twitter\footnote{\url{https://help.twitter.com/en/rules-and-policies/violent-groups}}, and YouTube\footnote{\url{https://support.google.com/youtube/answer/2801939?hl=en&ref_topic=9282436}}.

Here we define hate speech as \textit{episodes in which a speaker/user threatens, indulges, desires, or calls for physical violence against a target (e.g., minorities) or calls, denies or glorifies war crimes and crimes against humanity}. In other words, the notion of hate speech that we employ involves calls for violence against a target, in agreement with the literature and the regulators~\cite{siegel2019online}. 
Furthermore, we look at inappropriate (e.g. profanity) and offensive language (e.g. dehumanisation,  offensive remarks), which is not illegal, but deteriorates public discourse and can lead to a more radicalised society.

In this work, we analyse a corpus of more than one million comments on Italian YouTube videos related to COVID-19 to unveil the dynamics and trends of online hate. First, we manually annotate a large corpus of YouTube comments for hate speech and fine-tune a hate speech deep learning model to accurately detect it. Then, we apply the model to the entire corpus, aiming to characterise the behaviour of users producing hate, and shed light on the (possible) relationship between the consumption of misinformation and usage of hate and toxic language. 

This work advances the current literature at different levels. There is a large body of literature about community-level hate speech~\cite{kumar2018community,johnson2019hidden,mathew2020hate}. However, less is known about the behavioural features of users using hate speech on mainstream social media platforms, with few recent exceptions for Twitter~\cite{ribeiro2018characterizing,siegel2021trumping} and Gab~\cite{mathew2020hate}. Furthermore, to our knowledge, the relationship between online hate and misinformation is yet to be explored. In this paper, we study hate speech with respect to a controversial and heated topic, i.e., Covid-19, which has been already analysed in terms of sinophobic attitudes~\cite{schild2020go}. We relax the assumption behind many community-based studies, for which every post produced within an online community hosting haters is hate~\cite{johnson2019hidden,chandrasekharan2017bag}. 

Instead, to cope with a classification task that involves more than one million comments, we annotate a high-quality dataset of more than 70,000 YouTube comments, which is used for training a deep learning model. The model is standard in the state-of-the-art and builds on a wide strand of literature using machine learning~\cite{burnap2016us,del2017hate,davidson2017automated} and deep learning~\cite{badjatiya2017deep,basile2019semeval,zampieri2020semeval} for automatic classification of text and hate speech detection. Moreover, we distinguish YouTube channels into two categories: questionable, i.e., channels likely to disseminate misinformation, and reliable. This categorisation is in line with previous studies on the spreading of misinformation~\cite{cinelli2020covid,zollo2015emotional,zollo2017debunking}, and builds on a list of misinformation sources provided by the Italian Communications Regulatory Authority (AGCOM).

Our results show that hate speech on YouTube is slightly more present than on other social media platforms~\cite{gagliardone2016mechachal,siegel2021trumping} and that there are no significant differences between the proportions of hate speech detected in comments on videos from questionable and reliable channels. We also note that hate speech does not show specific temporal patterns, even on questionable channels. Interestingly, we do not find evidence of ``serial haters'', intended as active users posting exclusively hateful comments. Still, we note that users skewed towards one of the two categories of video channels (questionable, reliable) are more prone to use toxic language --i.e. inappropriate, violent, or hateful-- within their opponents community. Interestingly, users skewed towards reliable content use on average a more toxic language than their counterpart. Finally, we find that the overall toxicity of the discussion increases with its length measured both in terms of the number of comments and time. In other words, online debates tend to degenerate towards increasingly toxic exchanges of views, in line with Godwin's law.

\section{Methods}

\subsection{Data Collection}

We collected about 1.3M comments posted by more than 345,000 users on 30,000 videos from 7,000 channels on YouTube. 
Using the official YouTube Data API, we performed a keyword search for videos that matched a list of keywords, i.e., \{\textit{coronavirus, nCov, corona virus, corona-virus, covid, SARS-CoV}\}. An in-depth search was then performed by crawling the network of related videos as provided by the YouTube algorithm. Then, we filtered the videos that matched our set of keywords in the title or description from the gathered collection. Finally, we collected the comments received by these videos. The title and the description of each video, as well as the comments, are in Italian according to the Google's cld3 language detection service. The set of videos covers the time window that goes from 01/12/2019 to 21/04/2020, while the set of comments ranges in the time window that goes from 15/01/2020 to 15/06/2020.
 
We assigned a binary label to each YouTube channel to distinguish between two categories: questionable and reliable. A questionable YouTube channel is a channel producing unverified and false content or directly associated to a news outlet that failed multiple fact checks performed by independent fact checking agencies. The list of YouTube channels labelled as questionable 
was provided by the Italian Communications Regulatory Authority (AGCOM). The remainder of the channels were labelled as reliable. Table~\ref{tab:breakdown} shows a breakdown of the dataset.

\begin{table}[ht]
\centering
\begin{tabular}{c|c|ccc}
     &          & Channels & Videos  & Comments \\ \hline
\multirow{4}{*}{Category} & \multirow{2}{*}{Reliable}     & 7,140     & 29,975   & 1,170,461  \\
                          &                               & 99.7 \%  & 98.5 \% & 91.8 \%  \\
                          & \multirow{2}{*}{Questionable} & 17       & 464     & 103,475   \\
                          &                               & 0.3 \%   & 1.5 \%  & 8.2 \%   \\ \hline
\multirow{2}{*}{Total}    &          & 7,157     & 30,436   & 1,273,930  \\
                          &          & 100 \%   & 100 \%  & 100 \%  
\end{tabular}
\caption{Breakdown of YouTube data.}
\label{tab:breakdown}
\end{table}

\subsection{Hate Speech Model} 
Our aim is to create two high-quality manually annotated datasets for training and evaluating a deep learning hate speech model. We then apply the model to all the collected data and study the relationship between the hate speech phenomenon and misinformation.

Deep learning models based on Transformer architecture outperform other approaches to automated hate speech detection, as resulted from recent shared tasks in the SemEval-2019 evaluation campaign: HatEval~\cite{basile2019semeval} and OffensEval~\cite{zampieri-etal-2019-semeval}, as well as OffensEval 2020~\cite{zampieri2020semeval}.
The central reference for hate speech detection for Italian is the report on the EVALITA 2018 hate speech detection task~\cite{bosco2018overview}. 
Furthermore, in \cite{polignano2019hate} authors modelled the hate speech task using the Italian pre-trained language model AlBERTo achieving state-of-the-art results on Facebook and Twitter datasets. 
We trained a new hate speech detection model for Italian following the state-of-the-art approach \cite{polignano2019hate} on our four-class hate speech detection task 
(see Sections \ref{sub:data_selection} and \ref{sub:classification} for detailed information).

\subsubsection{Data Selection and Annotation}
\label{sub:data_selection}
The comments to be annotated were sampled from the Italian YouTube comments on videos about the COVID-19 pandemic in the period from January 2020 to May 2020. Two sets were annotated: a hate-speech-rich training set with 59,870 comments and an unbiased evaluation set with 10,536 comments.

To get a \textit{training set} that is rich with hate speech, we annotated all the comments with a (basic) hate speech classifier (machine learning model) that assigns a score between -3 (hateful) and +3 (normal). The basic classifier was trained on a publicly available dataset of Italian hate speech against immigrants~\cite{sanguinetti2018italian}. Even though this basic model is not very accurate, its performance is better than random and we used its result for selecting the training data to be annotated and later used for training our machine learning model. For a realistic evaluation scenario, threads (i.e. all the comments on a video) were kept intact during the annotation procedure, yet individual comments were annotated. 

The threads (with comments) to be annotated for the training set were selected according to the following criteria: thread length (the number of comments in a thread between ten and five hundred), and hatefulness (at least 5\% of hateful comments according to our basic classifier). The application of these criteria resulted in 1,168 threads (VideoIds) and 59,870 comments. 
The \textit{evaluation set} was selected from May 2020 data as a random (unbiased) sample of 10,543 comments grouped into 151 threads (videos).

Our annotation schema is adapted from OLID~\cite{zampierietal2019} and FRENK~\cite{ljubesic2019frenk}. We differentiate between the following hate speech types:
\begin{itemize}
     \item Appropriate; 
     \item Inappropriate (the comment contains terms that are obscene, vulgar; but the text is not directed at any person specifically); 
     \item Offensive (the comment includes offensive generalisation, contempt, dehumanisation, indirect offensive remarks); 
     \item Violent (the comment's author threatens, indulges, desires or calls for (physical) violence against a target; it also includes calling for, denying or glorifying war crimes and crimes against humanity). 
\end{itemize}

The data was split among eight contracted annotators. Each comment was annotated twice by two different annotators. The splitting procedure was optimised to get approximately equal overlap (in the number of comments) between each pair of annotators for each dataset. The  annotators were given clear annotation guidelines, a training session and a test on a small set to evaluate their understanding of the task and their commitment before starting the annotation procedure. Furthermore, the annotation progress was closely monitored in terms of annotator agreement to ensure high data quality.

\begin{figure}
     \centering
     \subfloat[][Training set]{\includegraphics[scale=0.44]{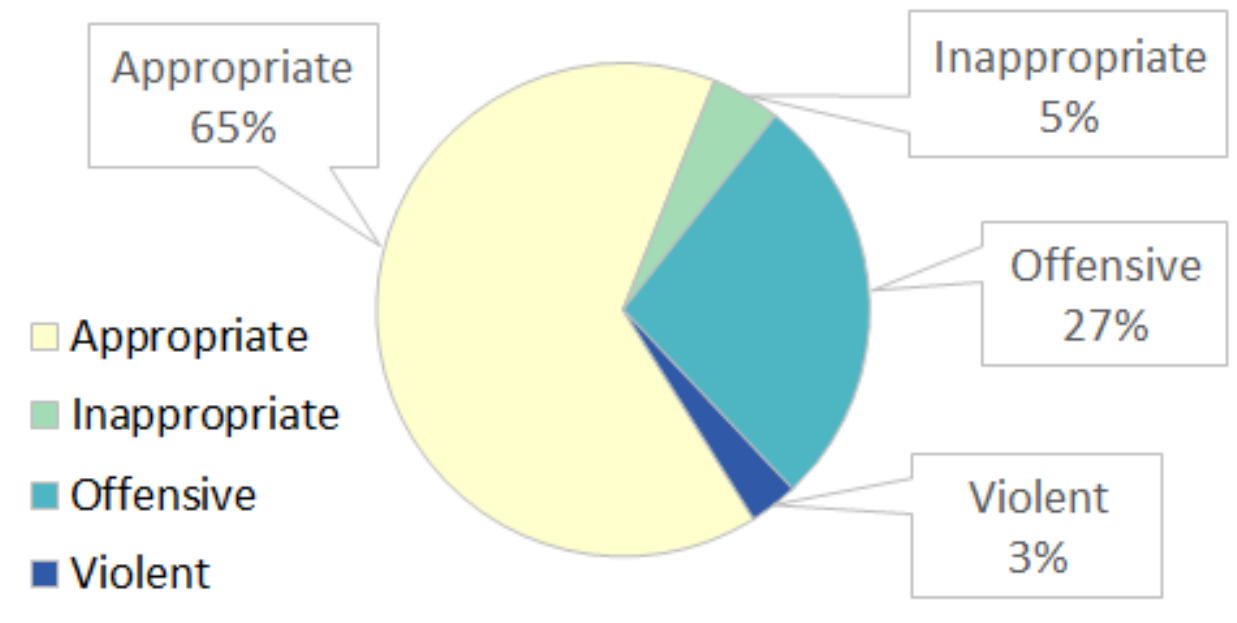}}
     \subfloat[][Evaluation set]{\includegraphics[scale=0.44]{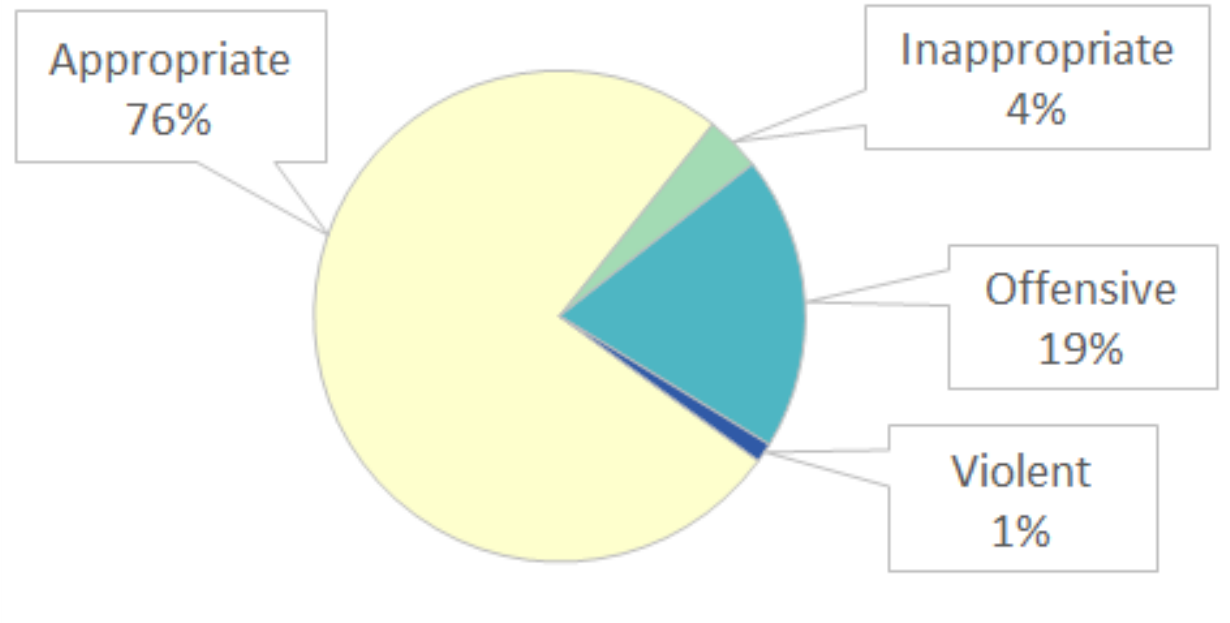}}
     \caption{The distribution of the four hate speech labels in the manually annotated training (a) and evaluation (b) sets. The training set is intentionally biased to contain more hate speech while the evaluation set is unbiased.}
     \label{fig:label_distribution}
\end{figure}

The annotation results for the training and evaluation sets are summarised in Figure~\ref{fig:label_distribution}. The annotator agreement in terms of Krippendorf alpha and accuracy (i.e. percentage of agreement) on both the training and the evaluation sets is presented in Table~\ref{tab:annotated_data}.
The agreement results indicate that the annotation task is difficult and ambiguous, as the annotators agree on the label of about 80\% of the cases.
Since we have an uneven class distribution, \alfa\, is a better measure of agreement as it accounts for the agreement by chance. 
Our agreement scores in terms of \alfa\, are comparable to those of high-quality datasets, like \cite{mozetic2016multilingual}.

\begin{table}[ht]
    \centering
    \caption{The annotation results for the  training and evaluation sets: Dataset size, date range, the annotator agreement in terms of interval Krippendorf alpha and accuracy (i.e. percentage of agreement).}
        \begin{tabular}{l|c|r|cccc}
    \hline
             
     Dataset & Date       & Size & \alfa & \acc \\
    \hline
        Train & Jan.-Apr. 2020 &119,670 & 0.586  & 0.774\\
        Evaluation & May 2020& 21,072 &   0.555 & 0.818  
    \end{tabular}
    \label{tab:annotated_data}
\end{table}

\subsubsection{Classification}
\label{sub:classification}
A state-of-the-art neural model based on Transformer language models was trained to distinguish between the four hate speech labels. We use a language model based on the BERT architecture \cite{devlin2018bert} which consists of 12 stacked Transformer blocks with 12 attention heads each. We attach a linear layer with a softmax activation function at the output of these layers to serve as the classification layer. As input to the classifier, we take the representation of the special [CLS] token from the last layer of the language model. The whole model is jointly trained on the downstream task of four-class hate speech detection. 
We used AlBERTo \cite{polignano2019alberto}, a BERT-based language model pre-trained on a collection of Tweets in the Italian language. According to previous work\cite{devlin2018bert}, fine-tuning of the neural models was performed end-to-end. We used the Adam optimizer with the learning rate of $2e-5$ and learning rate warmup over the first 10\% of the training instances. We used weight decay set to 0.01 for regularization. The model was trained for 3 epochs with batch size 32. We performed the training of the models using the HuggingFace Transformers library~\cite{Wolf2019HuggingFacesTS}.

The tuning of our models was performed by cross validation on the training set and the final evaluation was performed on the separate out-of-sample evaluation set.
In our setup, each data instance (comment) is labelled twice, possibly with inconsistent labels. To avoid data leakage between training and testing splits in cross validation, we use 8-fold cross validation where in each fold we use all comments annotated by one annotator as a test set.
We report the performance of the trained models using the same measures as we used for annotator agreement: 
 Krippendorff's Alpha-reliability (\alfa), accuracy (\acc), and add
 \fone\, score on both the training and the evaluation datasets. 
The validation results are reported in Table~\ref{tab:eval}.

\begin{table}%[ht]
\caption{Performance of our hate speech classification model on the training dataset (cross validation results) and the out-of-sample evaluation set.}
\label{tab:eval}
    \centering
    \small\addtolength{\tabcolsep}{-1pt}
    \begin{tabular}{r|cc|cccc}
    \hline
     & \multicolumn{2}{|c|}{overall} & appropriate & inappropriate & offensive & violent \\
     Model & \alfa & \acc & \fone & \fone & \fone & \fone  \\
    \hline
     Model train.   & 0.59 & 0.79 & 0.87 & 0.54 & 0.64 & 0.52   \\
     Model eval.   & 0.55 & 0.84 & 0.91 & 0.59 & 0.58 & 0.39   \\
    
    \hline
    An. agree. train.   & 0.59 & 0.77 & 0.86 & 0.52 & 0.63 & 0.63   \\
    An. agree. eval.  & 0.56 & 0.82 & 0.90 & 0.53 & 0.57 & 0.55   \\
    \hline
    \end{tabular}
\end{table}

The performance of our model is comparable to the annotator agreement in terms of Krippendorf Alpha and Accuracy, proving our model's high-quality. The model achieves the annotator agreement both on the training dataset in the cross validation setting as well as on the evaluation dataset which shows its ability to generalise well on the evaluation set. We observe similar results in terms of F1 scores for individual classes. The only noticeable drop in performance compared to the annotators is the performance on the minority (violent) class. We attribute this drop to the very low amount of data available for the violent class compared to the other classes, however the performance is still reasonable. We therefore apply our hate speech detection model to the set of 1.3M comments and report the findings.

\section{Results and Discussion}

\subsection{Relationship between hate speech and misinformation}

\begin{figure}
    \centering
    \includegraphics[scale = 0.4]{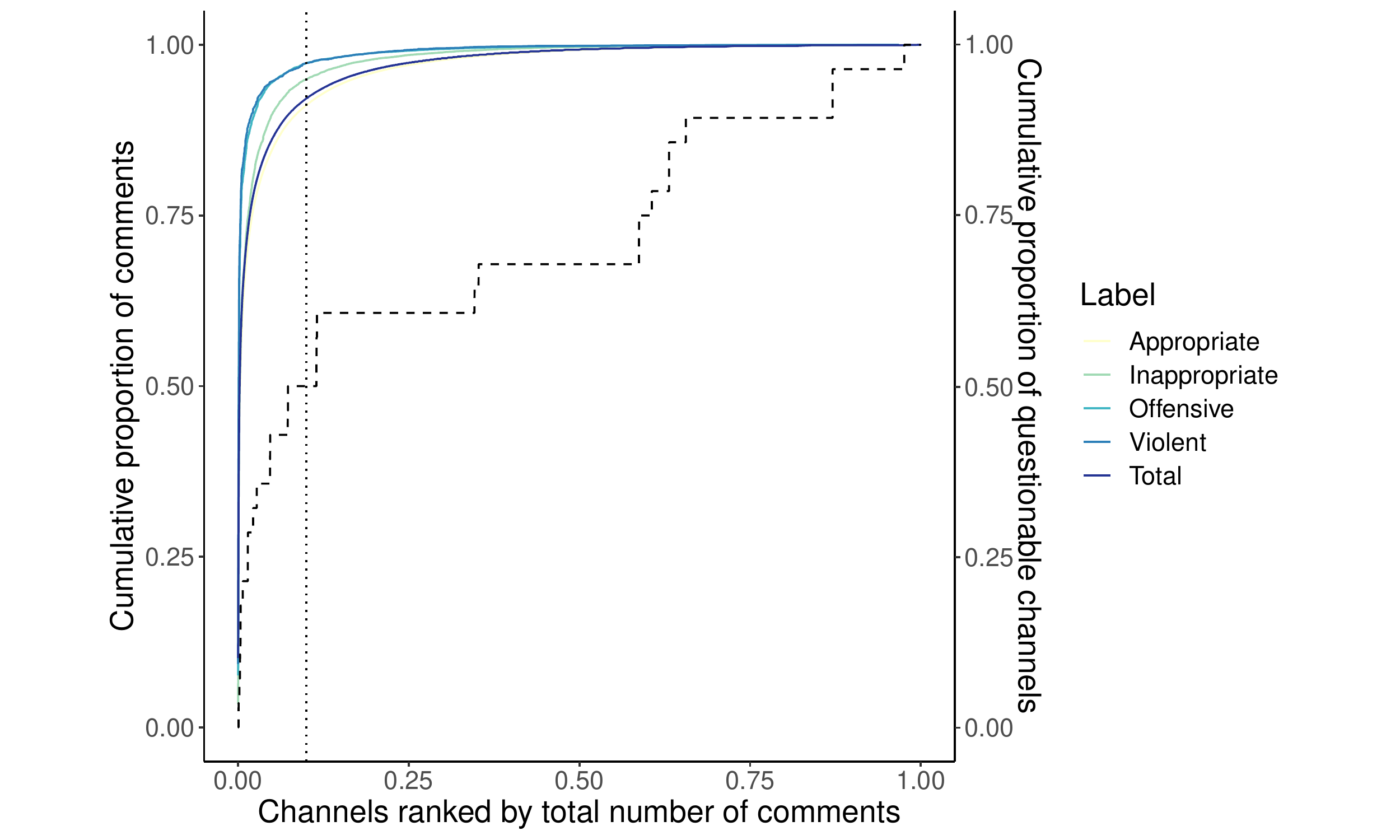}
    \caption{Ranking of YouTube channels by number of comments and proportions of comment types per channel.}
    \label{fig:Descriptive_Stats_1}
\end{figure}

We start our analysis examining the distribution of the different hate speech types on both reliable and questionable YouTube channels. Figure~\ref{fig:Descriptive_Stats_1} shows the cumulative distribution of comments, total and per type, by channel. The x-axis shows the YouTube channels ranked by their total number of comments, while the y-axis shows the total number of comments in the dataset (both quantities are reported as proportions). We observe that the distribution of comments is Pareto-like; indeed, the first 10\% of channels (dotted vertical line) covers about 90\% of the total number of comments. Such a 10 to 90 percent relationship is even stronger when comments are analysed according to their types; indeed, the heterogeneity of the distribution decreases going from violent to appropriate comments. It is also worth noting that, as indicated by the secondary y-axis of Figure~\ref{fig:Descriptive_Stats_1}, the first 10\% of channels with most comments also contain about 50\% of all the questionable channels in our list, thus indicating a relatively high popularity of these channels. In addition, questionable channels are about 0.25\% of the total number of channels that received at least one comment and, despite being such a minority, they cover $\sim$~8\% of the total number of comments (with the following partitioning: 8\% appropriate; 7\% inappropriate; 9\% offensive; 9\% violent) and the 1.3\% of the total number of videos, thus highlighting a disproportion between their activity and popularity.

\begin{figure}
    \centering
    \includegraphics[scale = 0.55]{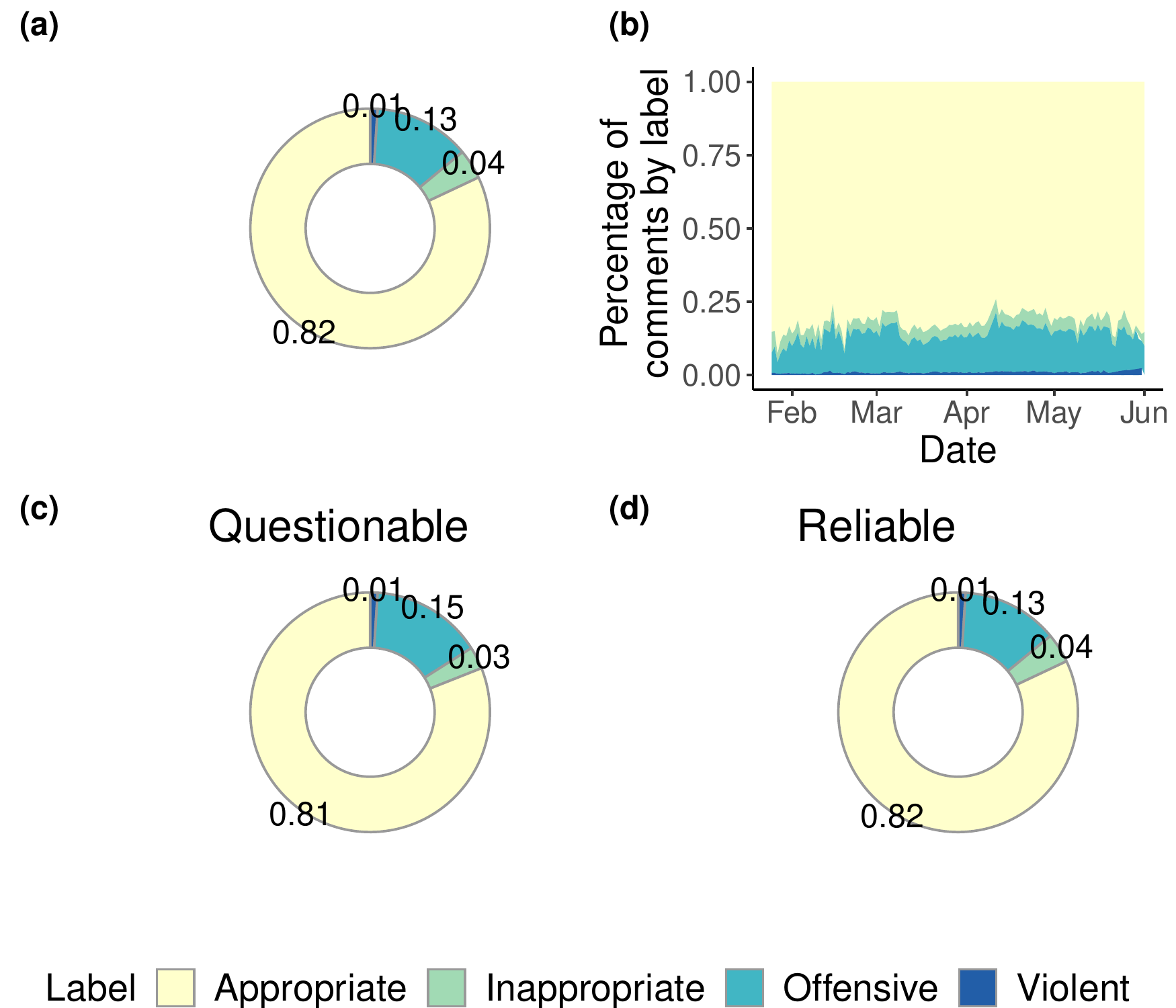}
    \caption{Proportion of the four hate speech labels in the whole dataset (a) over time (b), and for questionable (c) and reliable (d) YouTube channels.}
    \label{fig:Descriptive_Stats_2}
\end{figure}

Figure~\ref{fig:Descriptive_Stats_2} shows the proportion of comments by label and channel types, and their trend over time. In panel (a) we display the overall proportion of comment types, noting that the majority of comments is appropriate, followed by offensive, inappropriate, and violent types, all relatively stable over time (see panel (b)). It is worth remarking that, despite the proportion of hate speech found in the dataset is consistent with --although slightly higher than-- previous studies~\cite{gagliardone2016mechachal,siegel2021trumping}, the presence of even a limited number of hateful comments is in direct conflict with the platform's policy against hate speech. Moreover, we do not observe relevant differences between questionable (panel (c)) and reliable (panel (d)) channels, providing a first piece of evidence in favour of a moderate (if not absent) relationship between online hate and misinformation.

\begin{figure}
    \centering
    \includegraphics[scale = 0.35]{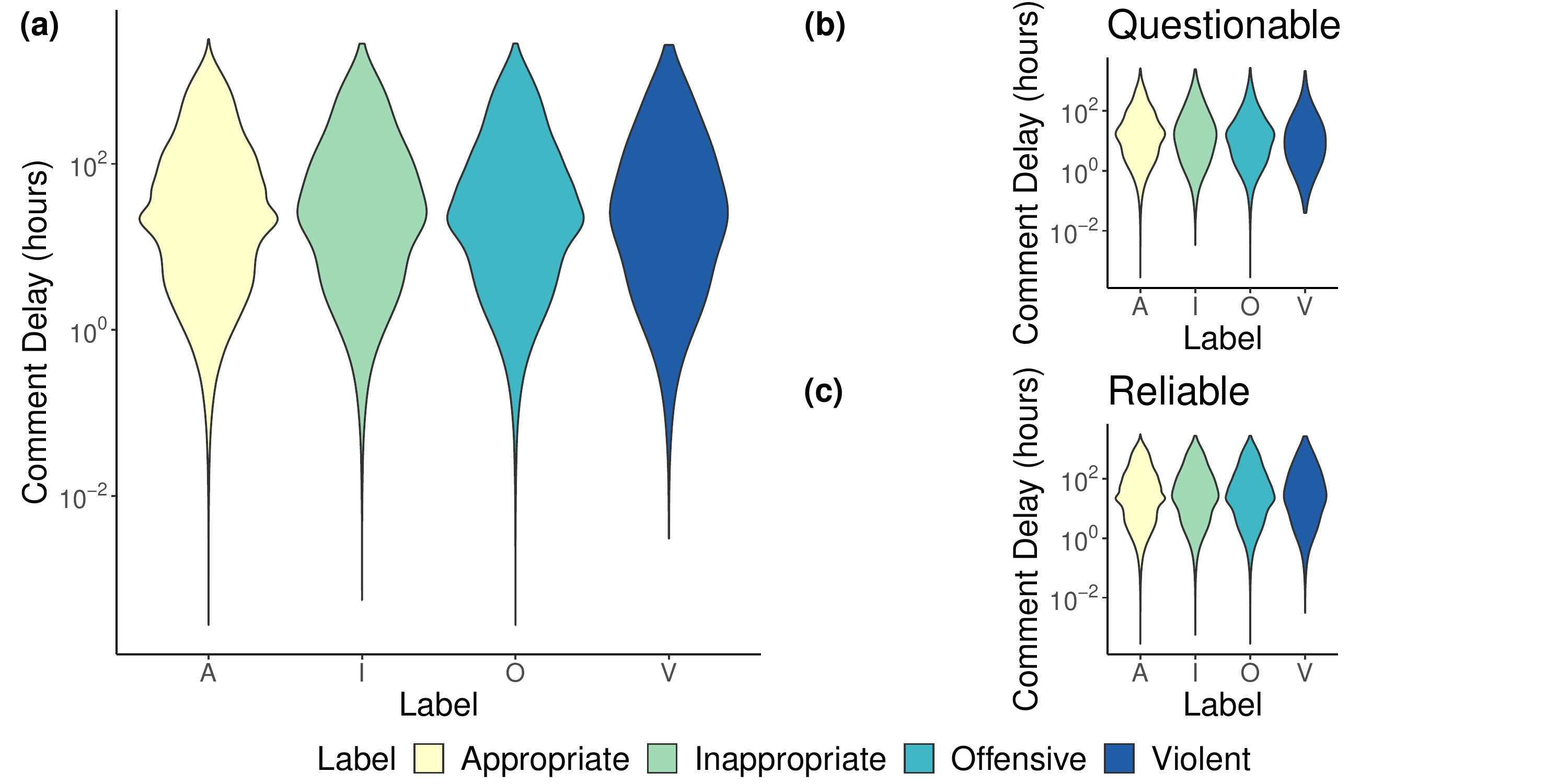}
    \caption{Distribution of comment delays in the whole dataset (a) and for questionable (b) and reliable (c) YouTube channels. The capital letters on the x-axis represent the different types of comments: appropriate (A); inappropriate (I); offensive (O); violent (V).}
    \label{fig:delay}
\end{figure}

Now we aim at understanding whether hateful comments display a typical (technically, the average) time of appearance. This kind of information can indeed be crucial for the implementation of timely moderation efforts. More specifically, our goal is to discover whether 1) different hate speech types have typical delays and 2) any difference holds between comments on videos disseminated by questionable and reliable channels.
To this aim, we define the comment delay as the time elapsed between the posting time of the video and that of the comment (in hours). Figure~\ref{fig:delay} displays the comment delays for the four types of hate speech and for questionable and reliable channels. Looking at panel (a) of Figure~\ref{fig:delay}, we first note that all comments share approximately the same delay regardless of their type. Indeed, the distributions of the comment delay are roughly log-normal with a long average delay ranging from 120 hours in the case of appropriate comments to 128 hours in the case of violent comments (the comment delay is reduced by $\sim75\%$ when removing observations in the right tail of the distribution as shown in Table~\ref{tab:delay_general} of SI). For what concerns comments on videos published by questionable and reliable channels, we do not find strong differences between typical delays of hate speech types within the two domains. In the case of questionable channels, we find that comment delays range from 66 to 42 hours, while for reliable channels they range from 125 to 136 hours (as reported in SI). To summarise, we find a discrepancy in users' responsiveness to the two types of content, with comments on questionable videos having a much lower typical delay than those on reliable videos. In addition, comments typical delays differ between reliable and questionable channels. In particular, on questionable channels toxic comments appear first and faster than appropriate ones, following decreasing levels of toxicity (violent $\rightarrow$ offensive $\rightarrow$ inappropriate). In other words, violent comments on questionable content display the shortest typical delay, followed by offensive, inappropriate, and appropriate comments. Conversely, on reliable channels the shortest typical delay is observed for appropriate comments, followed by violent, inappropriate, and offensive comments (for details refer to SI).

\subsection{Users' behaviour and misinformation}

\begin{figure}
    \centering
    \includegraphics[scale = 0.5]{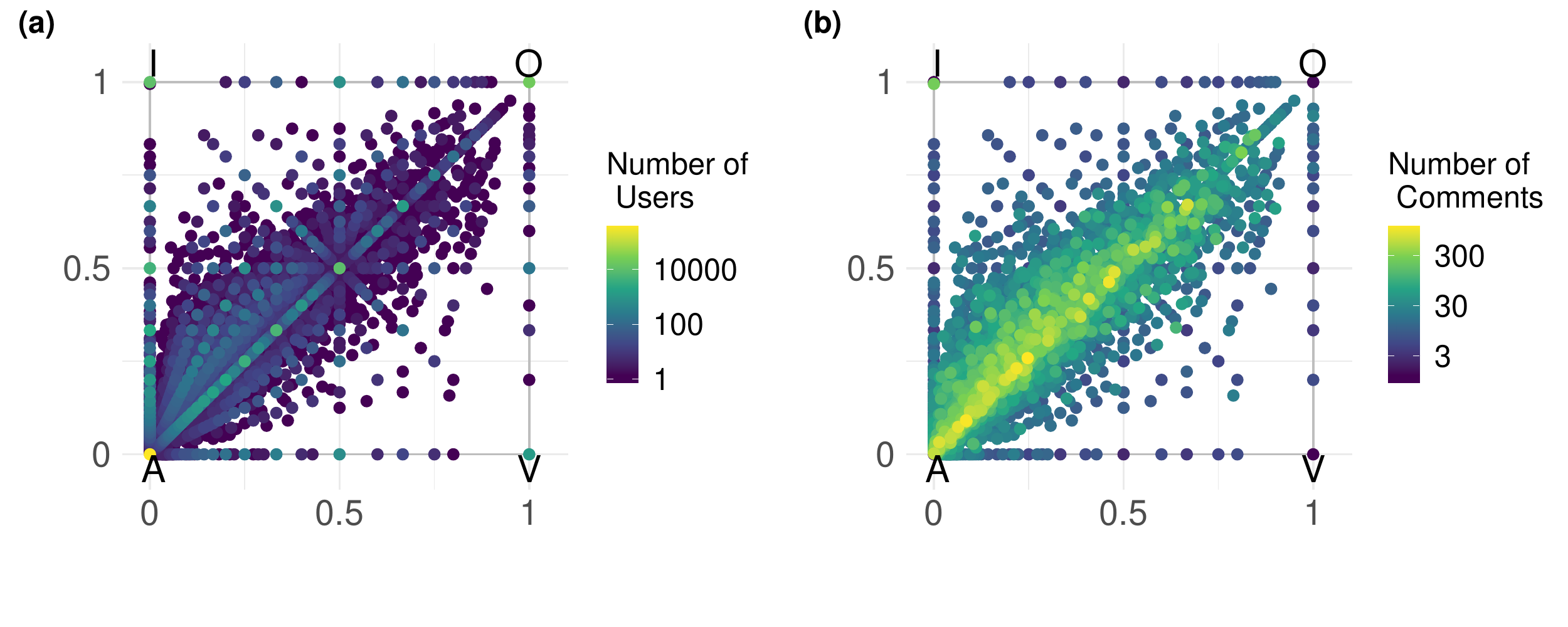}
    \caption{Users balance between different comment types. In panel (a) brighter dots indicate a higher density of users while in panel (b) brighter dots indicate a higher average activity of the users in terms of number of comments. We note that users focused on posting comments labelled as offensive and violent are almost absent in the data.} 
    \label{fig:no_haters_1}
\end{figure}

In line with other social media platforms~\cite{del2016spreading, cinelli2020covid}, users activity on YouTube follows a heavy tailed distribution, i.e., the majority of users post few comments, while a small minority is hyperactive (see SI for details). Now we want to investigate whether a systematic tendency towards offences and hate can be observed for some (category of) users. In Figure~\ref{fig:no_haters_1}, each vertex of the square represents one of the four hate speech types (appropriate - A; inappropriate - I; offensive - O; violent - V). Each dot is a user whose position in the square depends on the fraction of his/her comments for each category. As an example, a user posting only appropriate comments will be located exactly on the vertex A (i.e., in (0,0)), while a user that splits his/her activity evenly between appropriate and inappropriate comments will be located in the middle of the edge connecting the vertices A and I. Similarly, a user posting only violent comments will be located exactly on the vertex V (i.e., in (1,0). 
More formally, to shrink the 4-dimensional space deriving by the four labels that fully characterise the activity of each user, we associate a user $j$ the following coordinates in a 2-dimensional space:

\begin{equation}
    x_j = a_j*0 + i_j*0 + o_j*1 + v_j*1
\end{equation}
\begin{equation}
    y_j = a_j*0 + i_j*1 + o_j*1 + v_j*0 
\end{equation}

\noindent where $a_j$, $i_j$, $o_j$, $v_j$ are the proportions, respectively, of appropriate, inappropriate, offensive, and violent comments posted by user $j$ over his/her total activity $c_j$.

%****************
%no serial haters
%****************
Although most of the users leave only or mostly appropriate comments, there are also several users ranging across categories (i.e., located away from the vertices of the square in Figure~\ref{fig:no_haters_1}). Interestingly, there is no evidence of ``serial haters", i.e., active users exclusively using hateful language. Indeed, while there are users posting only or mostly violent comments (see Figure~\ref{fig:no_haters_1} a), their overall activity is very low and below five comments (see Figure~\ref{fig:no_haters_1} b). A similar situation is observed for offenders, i.e., active users posting only offending comments. Although we cannot exclude that moderation efforts put in place by YouTube (if any) might partially impact these results, the absence of serial haters and offenders highlights that hate speech is rarely only an issue of specific categories of users. Rather, it seems that regular users are occasionally triggered by external factors.  

To rule out possible confounding factors (note that users located in the centre of the square could display a balanced activity between different pairs of comment categories) we repeated the analysis excluding the category I (i.e, inappropriate). The results are provided in SI and confirm what we observe in Figure~\ref{fig:no_haters_1}. 

We now aim at unveiling the relationship between users behaviour in terms of commenting patterns and their activity with respect to questionable and reliable channels. Since misinformation is often associated with the diffusion of polarising content which plays on one's fear and could fuel anger, frustration and hate ~\cite{delvicario2019polarization, osmundsen2020partisan, guess2019less}, our intent is to understand whether users more loyal to questionable content are also more prone to use a toxic language in their comments. 
Thus, we define the leaning $l$ of a user $j$ as the fraction of his/her activity spent in commenting videos posted by questionable channels, i.e., 

\begin{equation}
    l_j = \sum_{i = 1}^{c_j}\frac{q_j}{c_j}
\end{equation}

\noindent where $\sum_{i = 1}^{c_j}q_j$ is the number of comments on videos from questionable channels posted by the user $j$ and $c_j$ is the activity of user $j$. Similarly, for each user $j$ we compute the fraction of non-appropriate comments $\overline{a}$ as:

\begin{equation}
    \overline{a}_j = 1 - a_j
\end{equation}

\noindent where $a_j$ is the fraction of appropriate comments posted by user $j$.

\begin{figure}
    \centering
    \includegraphics[scale = 0.35]{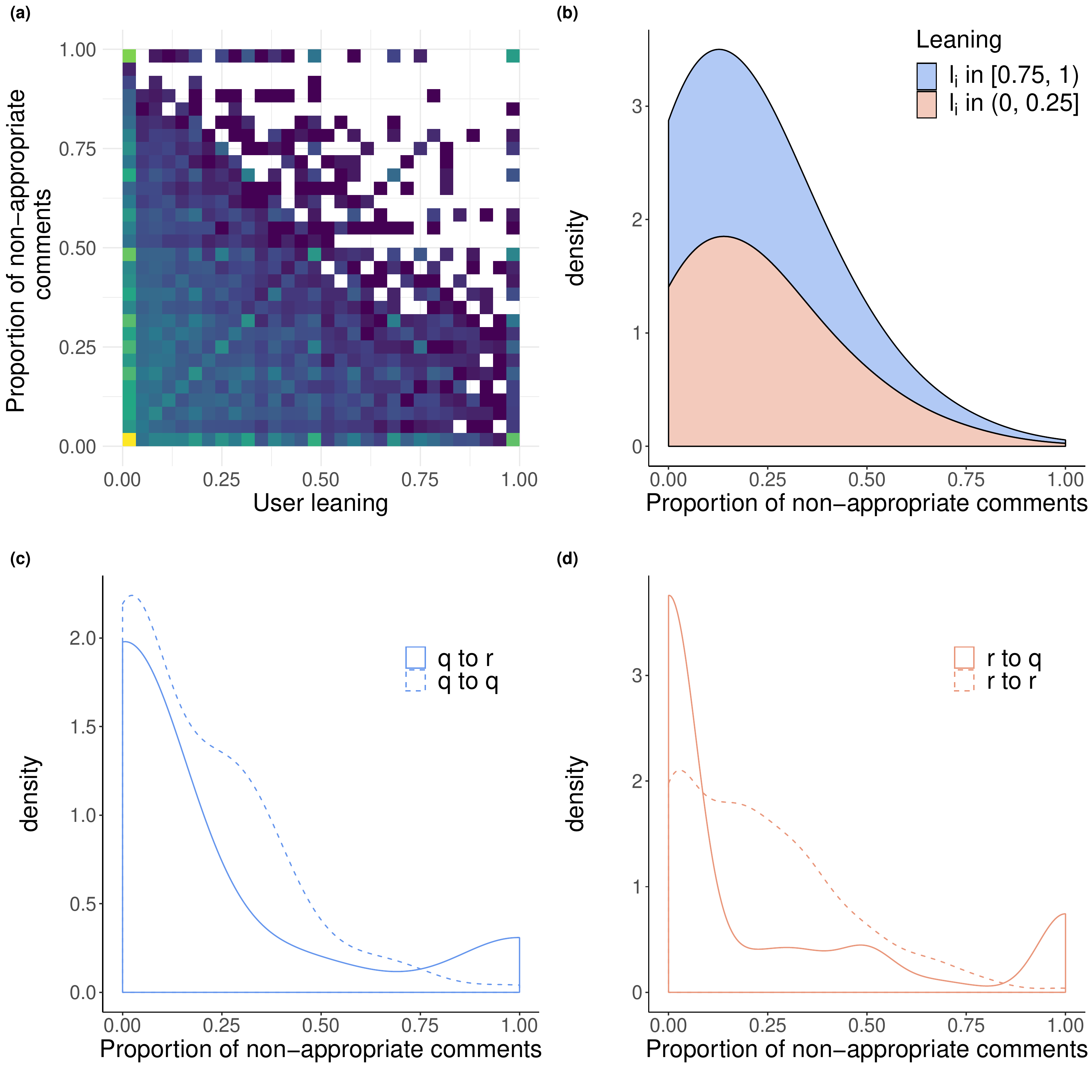}
    \caption{Panel (a) displays the relationship occurring between the preference of users for questionable and reliable channels (the user leaning $l_j$) and the fraction of non-appropriate comments posted by the user ($\overline{a}_j$) as a joint distribution. Panel (b) displays the distribution of non-appropriate comments for users displaying a remarkable tendency to comment under videos posted by questionable ($l_j \in [0.75,1)$) and reliable ($l_j \in (0,0.25]$) channels. Panel (c) displays the distribution of non-appropriate comments posted by users with leaning towards questionable channels ($l_j \in [0.75,1)$ indicated as q) under videos of questionable channels (dashed line indicated as q to q in the legend) and under videos of reliable channels (solid line indicated as q to r in the legend). Panel (d) displays the distribution of non-appropriate comments posted by users with leaning towards reliable channels ($l_j \in (0,0.25]$ indicated as r) under videos of questionable channels (solid line indicated as r to q in the legend) and under videos of reliable channels (dashed line indicated as r to r in the legend).}
    \label{fig:users_pol}
\end{figure}

In Figure~\ref{fig:users_pol} (a), we compare users' leaning $l_j$ against the fraction of non-appropriate comments $\overline{a}_j$. As expected, we may observe two peaks (of different magnitude) in correspondence of extreme values of leaning ($l_j \sim 0$ and $l_j \sim 1$), represented by the brighter squares in the plot. In addition, the joint distribution becomes sparser in correspondence of higher values of users' leaning and fraction of non-appropriate comments ($l_j \geq 0.5$ and $\overline{a}_j \geq 0.5$), indicating that a relevant share of users are placed at the two extremes of the distribution (thus being somewhat polarised) and that users producing mostly non-appropriate comments are way less present.

In Figure~\ref{fig:users_pol} (b), we display the proportion of non-appropriate comments posted by users displaying leaning at the two tails of the distribution (i.e., users displaying a remarkable tendency to comment questionable videos $l_j \in [0.75,1)$ and users with a remarkable tendency to comment reliable videos $l_j \in (0,0.25]$). We find that users skewed towards reliable channels post, on average, a higher proportion of non-appropriate comments ($\sim23\%$) than users skewed towards questionable channels ($\sim17\%$). In other words, users who tend to comment on reliable videos are also more prone to use a non-appropriate/toxic language. Further statistics on the two distributions are reported in SI.

Panel (c) of Figure~\ref{fig:users_pol} provides a comparison between the distributions of non-appropriate comments posted by users skewed towards questionable channels ($q$ in the legend) on videos published by either questionable or reliable channels. Panel (d) of Figure~\ref{fig:users_pol} provides a similar representation for users skewed towards reliable channels ($r$ in the legend). We may note a strong difference in users behaviour: quite unimodal when they comment videos on the same side of the leaning; bimodal when they comment videos on the opposite side of leaning. Therefore, users tend to avoid using a toxic language when they comment videos in accordance with their leaning and to separate into roughly two classes (non-toxic, toxic) when they comment videos in contrast with their preferences. This finding resonates with evidence of online polarisation and with the presence of peculiar characters of the internet such as trolls and social justice warriors.

%*****
%reliable users are more prone to be offensive
%*****

\subsection{Toxicity Level of Online Debates}
Finally, we aim at investigating whether online debates degenerate (i.e., increase their average toxicity) when the discussion gets longer, both in terms of number of comments and time. Indeed, we are interested in analysing how commenting dynamics change over time and whether online hate follows similar dynamics to those observed for users' sentiment~\cite{zollo2015emotional}. Moreover, we want to understand whether the toxicity of comments tends to follow certain dynamics empirically observed on the internet such as Godwin's law. To this purpose, we test whether toxic comments tend to appear more frequently at later stages of the debate. 

To compute the toxicity level of a debate around a certain video, we assign each hate speech type (A,I,O,V) a toxicity value $t$ as follows: 
\begin{itemize}
    \item Appropriate: $t$ = 0
    \item Inappropriate: $t$ = 1
    \item Offensive: $t$ = 2
    \item Violent: $t$ = 3
\end{itemize}
Then, we define the toxicity level $T$ of a discussion $d$ of $n$ comments as the average of the toxicity values over all the comments of the discussion:

$$T_d = \frac{\sum_{j=1}^{n}{t_j}}{n}$$

\begin{figure}
    \centering
    \includegraphics[scale = 0.58]{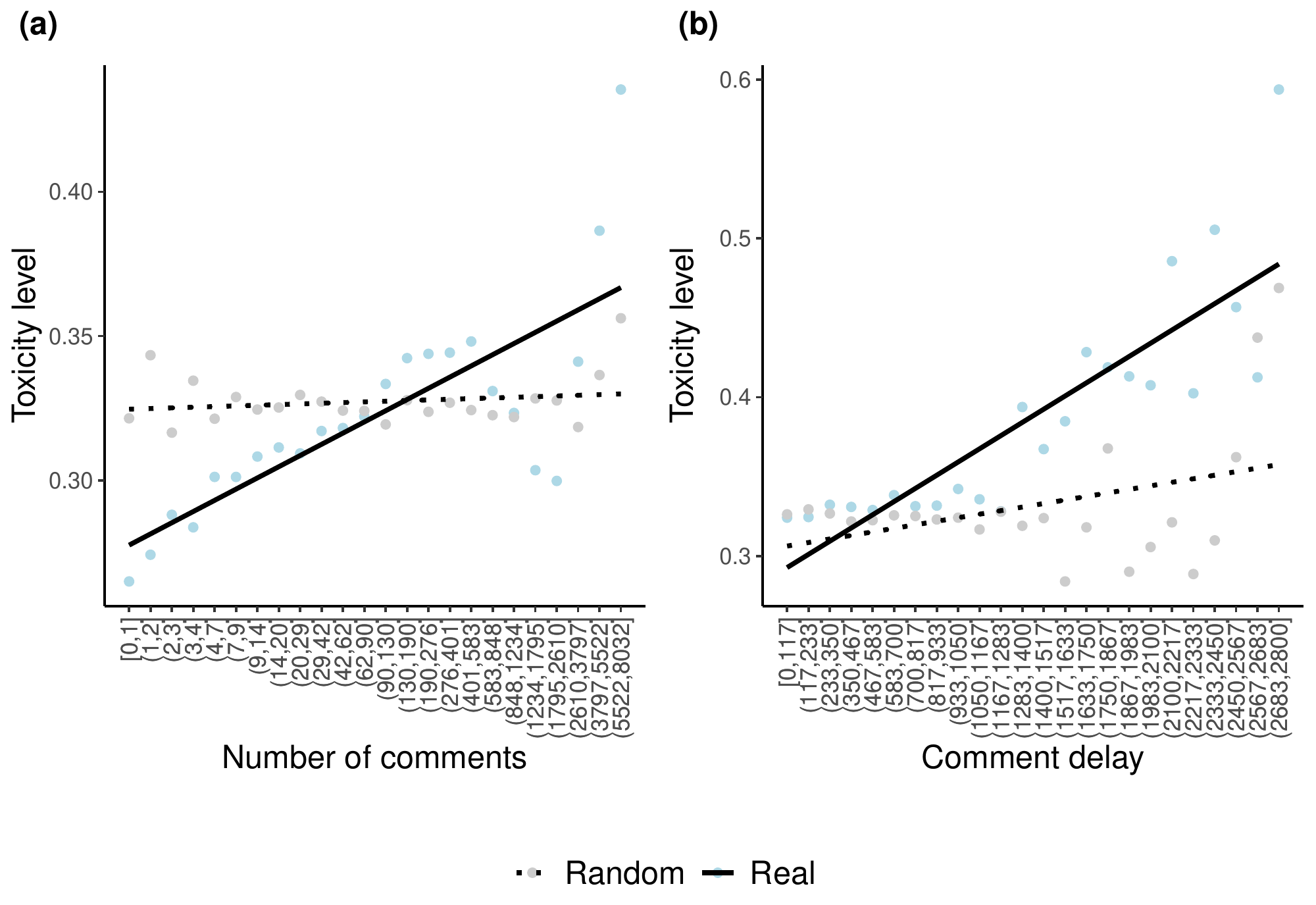}
    \caption{Linear regression models for number of comments and comment delay. On the x-axis of panel (a) the comments are grouped in logarithmic bins while on the x-axis of panel (b) the comment delays are grouped in linear bins.}
    \label{fig:toxicity}
\end{figure}

To understand how the toxicity level changes with respect to the number of comments and to comment delay (i.e., the time elapsed between the posting time of the video and that of the comment), we employ linear regression models. Figure~\ref{fig:toxicity} shows that a positive relationship between the two variables (i.e., average toxicity is an increasing function of the number of comments and comment delay) exists, and that such a relationship cannot be reproduced by linear models obtained with randomised comment labels (regression outcomes are reported in SI). 

We apply a similar approach to distinguish between comments on videos from questionable and reliable channels (as shown in SI). Overall, similarly to the general case, we find stronger positive effects in real data than in randomised models although such effects are significant only in the case of comments under videos posted by reliable channels. 

\section{Conclusions}
The aim of this work is two-fold: i) to investigate the behavioural dynamics of online hate speech and ii) to shed light on the possible relationship with misinformation exposure and consumption. We apply a hate speech deep learning model to a large corpus of more than one millions comments on Italian YouTube videos. 
Our analysis provides a series of important results which can support the development of appropriate solutions to prevent and counter the spread of hate speech online. 
\textit{First}, there is no evidence of a strict relationship between the usage of a toxic language (including hate speech) and being involved within the misinformation community on YouTube.
\textit{Second}, we do not observe the presence of ``serial'' haters, instead it seems that the phenomenon of hate speech involves regular users who are occasionally triggered to use toxic language.
\textit{Third}, users polarisation and hate speech seem to be intertwined, indeed users are more prone to use inappropriate, violent, or hateful language within their opponents community (i.e., out of their echo chamber).
\textit{Finally}, we find a positive correlation between the overall toxicity of the discussion and its length, measured both in terms of number of comments and time.

Our results are in line with recent studies about (the increasing) polarisation of online debates and segregation of users~\cite{cinelli2021echo}. Furthermore, they somewhat confirm the intuition behind some empirically grounded laws such as Godwin's law which can be interpreted, by extension, as a statement regarding the increasing toxicity of online debates.
A potential limitation of this work is represented by the relentless effort of YouTube in moderating hate on the platform. This could have prevented us from having complete information about the actual presence of hate speech in public discussions. 
Future efforts should extend our work to other languages beyond Italian, social media platforms, and topics. For instance, studying hate speech on online political discourse over time could provide important insights on debated phenomena such as affective polarisation\cite{druckman2021affective}. Moreover, further research on possible triggers in the language and content of videos is desirable.

\section*{Acknowledgements}
The authors acknowledge financial support from the Slovenian Research Agency (research core funding no. P2-103), and the European Union’s Rights, Equality and Citizenship Programme under Grant Agreement no. 875263. The authors wish to thank Arnaldo Santoro for his support with the categorisation of misinformation sources. 

\bibliographystyle{unsrt}
\bibliography{references}

\newpage

\section*{Supplementary Information}

\begin{table}[th]
    \centering
    \begin{tabular}{l|cccc}
                    & $\mu_{cd}$ & $\sigma_{cd}$ & $\gamma_{cd}$ & $20\% \mu_{cd}$ \\ \hline
       Appropriate  &  120 & 259 & 3.85 & 35.4\\
       Inappropriate & 128 & 269 & 3.77 & 39.7\\
       Offensive & 128 & 278 & 3.84 & 36.8\\
       Violent & 127 & 272 & 3.87 & 38.2\\
    \end{tabular}
    \caption{Comment delays in hours for each category. The first column displays the average comment delay ($\mu_{cd}$); the second column displays the standard deviation of the comment delay ($\sigma_{cd}$); the third column displays the skewness ($\gamma_{cd}$), i.e., the asymmetry of the comment delays that is always positive thus indicating the presence of deviations in the right tail; the fourth column displays the 20\% trimmed mean ($20\% \mu_{cd}$), i.e., the average comment delay after removing the 20\% of comments having the highest comment delays.}
    \label{tab:delay_general}
\end{table}

% \usepackage{graphicx}
\begin{table}[ht]
%\resizebox{\textwidth}{!}{%
\begin{tabular}{l|cccc}
              & \multicolumn{2}{c}{$\hat{\mu}_{cd}$} & \multicolumn{2}{c}{$\hat{\sigma}_{cd}$} \\
              & Questionable            & Reliable               & Questionable               & Reliable                \\ \hline
Appropriate   &  65.9          &  124.6         &  7.4              &  6.7              \\
Inappropriate &  65.5          &  132.2        & 7.3            & 7.1             \\
Offensive     & 53.3         & 135.0         & 9.1            &  6.9            \\
Violent     &  41.7          &  135.9        &  7.8             & 7.4            
\end{tabular}%
%}
\caption{Comment delays in hours for each category of comments and channels. Since the two categories have different sample size the summary statistics reported in this table derive from a bootstrap of 7,500 samples per category repeated 1,000 times. The first column displays the average bootstrap comment delay ($\hat{\mu}_{cd}$); the second column displays the standard deviation of the average bootstrap comment delay ($\hat{\sigma}_{cd}$).}
\label{tab:delay_qr}
\end{table}

\begin{figure}
    \centering
    \includegraphics[scale = 0.4]{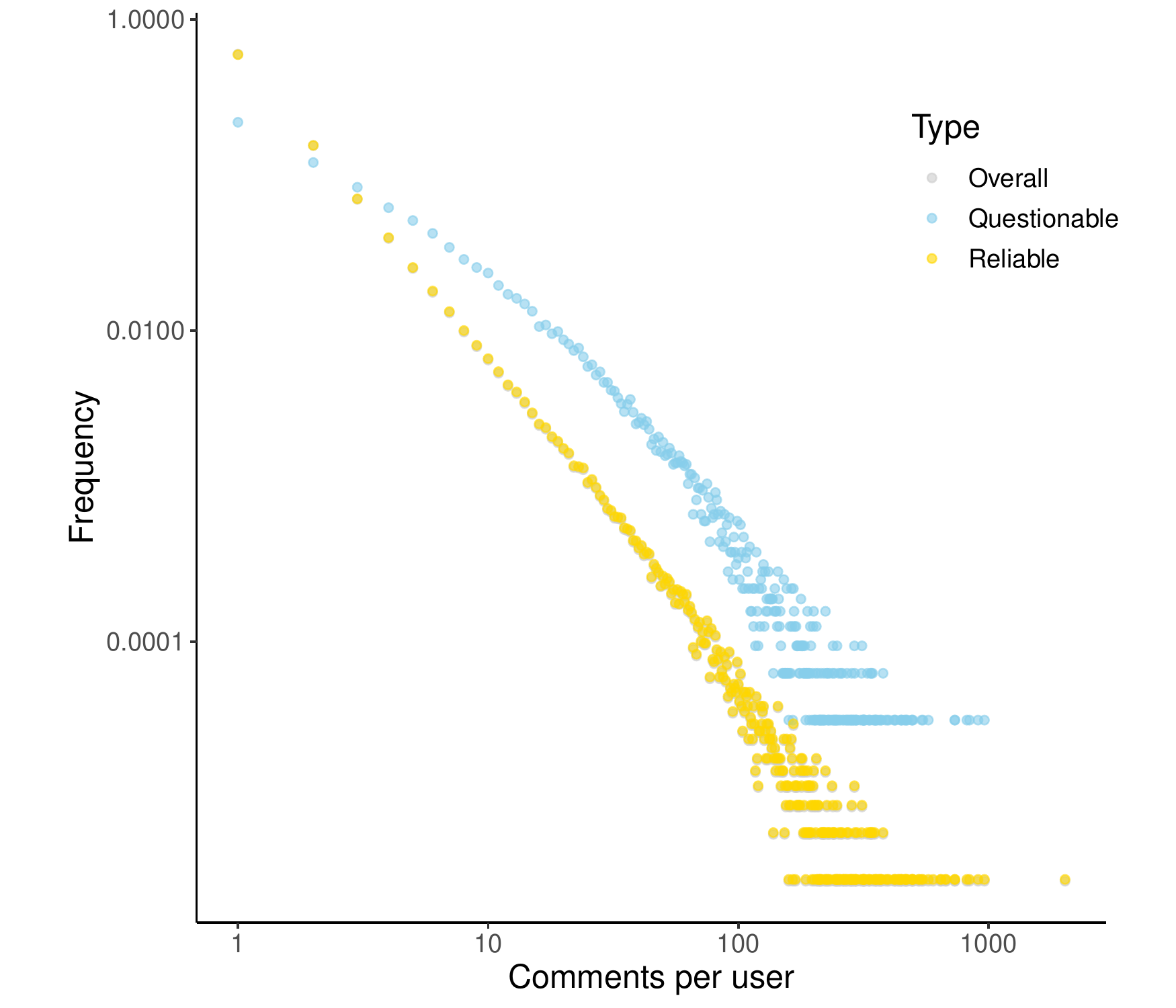}
    \caption{Distribution of comments per user in the general case and restricting the count only to comments posted by the user under either questionable or reliable videos.}
    \label{fig:user_activity}
\end{figure}

\begin{figure}
    \centering
    \includegraphics[scale=0.4]{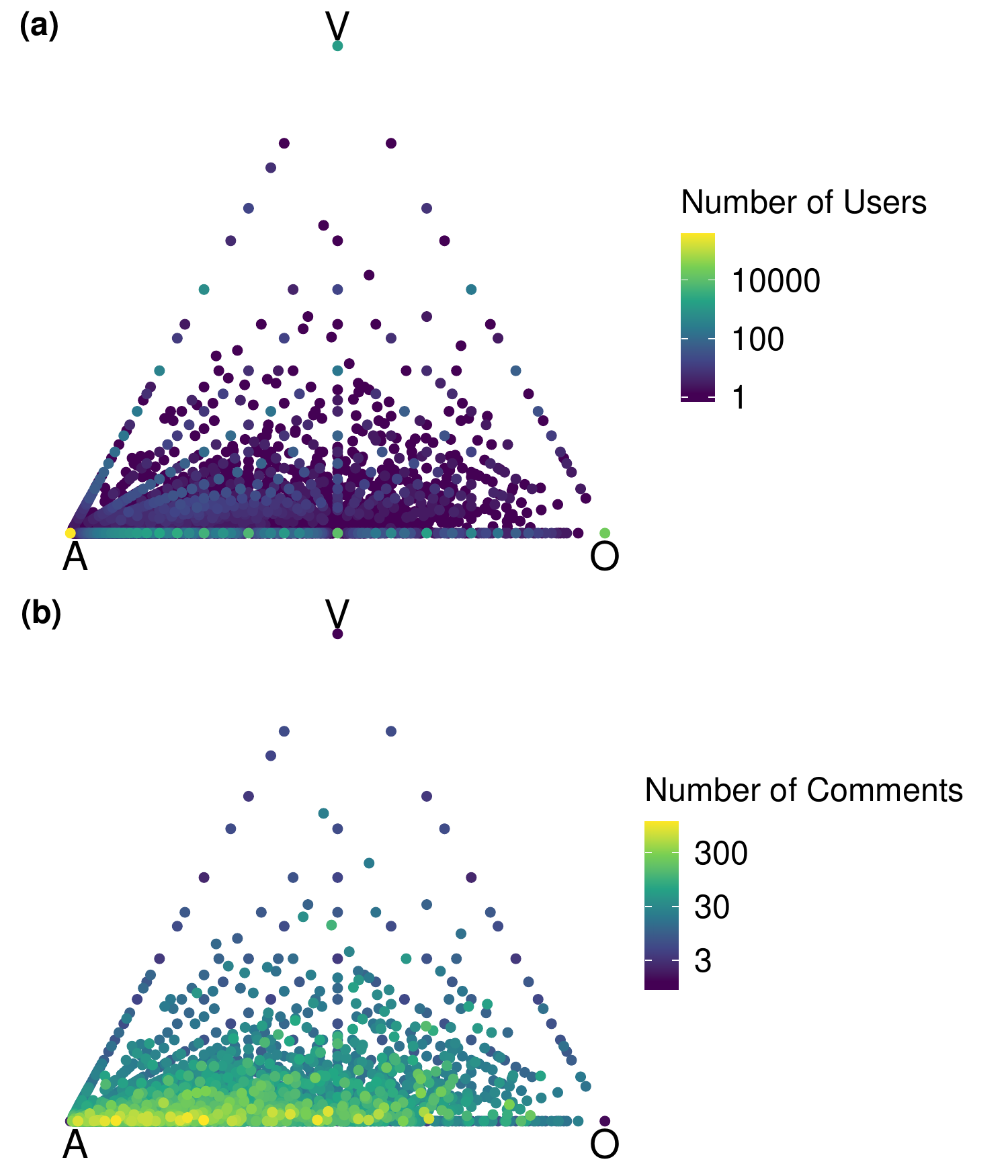}
    \caption{Users balance between different comment types. Comments labelled as inappropriate (I) are eliminated with respect to Figure~\ref{fig:no_haters_1}. Each dot in mapped into the triangle using barycentric coordinates. In panel (a) brighter dots indicate a higher density of users while in panel (b) brighter dots indicate a higher average activity of the users in terms of number of comments.  Consistently with Figure~\ref{fig:no_haters_1}, we note that users focused on posting comments labelled as offensive and violent are almost absent in the data.}
    \label{fig:no_serial_haters_2}
\end{figure}

\begin{table}[th]
    \centering
    \begin{tabular}{l|ccc}
                    & $\mu_{\overline{a}}$ & $\sigma_{\overline{a}}$ & $\gamma_{\overline{a}}$ \\ \hline
       $l_j \in (0, 0.25]$  &  0.23 & 0.19 & 0.94\\
       $l_j \in [0.75,1)$ & 0.17 & 0.20 & 1.29\\
    \end{tabular}
    \caption{Summary statistics of proportions of non-appropriate comments by users with leaning skewed towards reliable ($l_j \in (0, 0.25]$) and questionable ($l_j \in [0.75,1)$) channels. The first column displays the average proportion of non-appropriate comments ($\mu_{\overline{a}}$); the second column displays the standard deviation of the proportion of non-appropriate comments ($\sigma_{\overline{a}}$); the third column displays the skewness ($\gamma_{\overline{a}}$), i.e., the asymmetry of the distributions, that is positive thus indicating the presence of deviations in the right tail.}
    \label{tab:fig6_b}
\end{table}

\begin{table}
\begin{center}
\small
\begin{tabular}{l c c c c}
\hline
 & Model 1 & Model 2 & Model 3 & Model 4 \\
 & Number of comments & Number of comments & Comment delay  & Comment delay \\
 & (Real) & (Random) & (Real) & (Random) \\
\hline
(Intercept) & $0.2736^{***}$ & $0.3244^{***}$ & $0.2844^{***}$ & $0.3040^{***}$ \\
            & $(0.0101)$     & $(0.0036)$     & $(0.0158)$     & $(0.0167)$     \\
x           & $0.0039^{***}$ & $0.0002$       & $0.0083^{***}$ & $0.0022$       \\
            & $(0.0007)$     & $(0.0003)$     & $(0.0011)$     & $(0.0012)$     \\
\hline
R$^2$       & $0.5768$       & $0.0365$       & $0.7188$       & $0.1414$       \\
Adj. R$^2$  & $0.5576$       & $-0.0073$      & $0.7061$       & $0.1024$       \\
%Num. obs.   & $24$           & $24$           & $24$           & $24$           \\
\hline
\multicolumn{5}{l}{\scriptsize{$^{***}p<0.001$; $^{**}p<0.01$; $^{*}p<0.05$}}
\end{tabular}
\caption{Statistical models of Figure~\ref{fig:toxicity} a and b.}
\label{tab:reg_coefficients}
\end{center}
\end{table}

\begin{figure}
    \centering
    \includegraphics[scale=0.4]{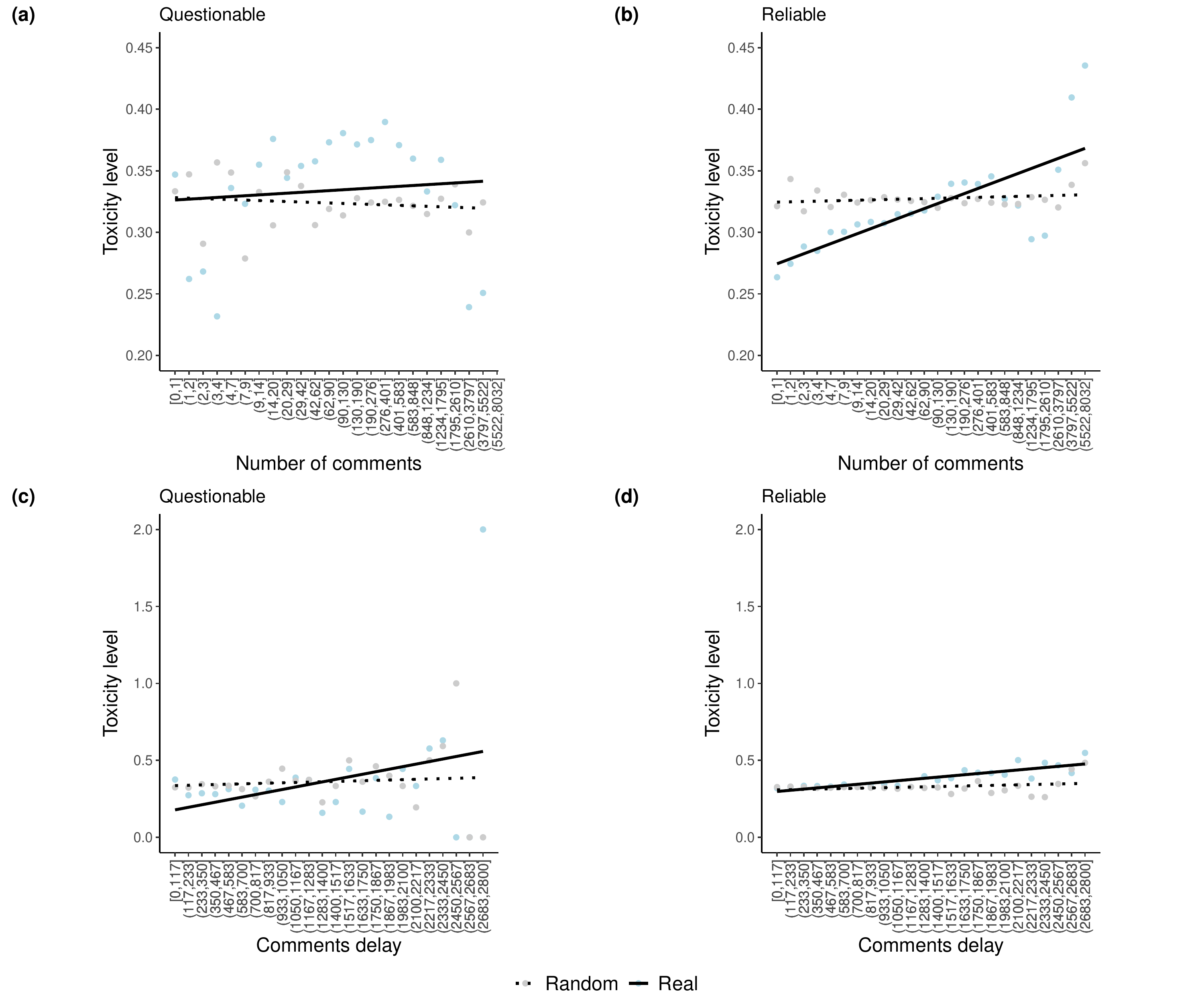}
    \caption{Linear regression models for number of comments and comment delay separating videos posted by questionable (panels (a) and (c)) and reliable channels (panels (b) and (d)). On the x-axis of panel (a) and (b) the comments are grouped in logarithmic bins while on the x-axis of panel (c) and (d) the comment delays are grouped in linear bins.}
    \label{fig:toxicity_qr}
\end{figure}

\begin{table}
\begin{center}
\begin{tabular}{l c c c c}
\hline
 & Model 1 & Model 2 & Model 3 & Model 4 \\
 & Questionable & Questionable & Reliable  & Reliable \\
 & (Real) & (Random) & (Real) & (Random) \\
\hline
(Intercept) & $0.3256^{***}$ & $0.3285^{***}$ & $0.2704^{***}$ & $0.3243^{***}$ \\
            & $(0.0214)$     & $(0.0084)$     & $(0.0110)$     & $(0.0036)$     \\
x           & $0.0007$       & $-0.0004$      & $0.0041^{***}$ & $0.0003$       \\
            & $(0.0016)$     & $(0.0006)$     & $(0.0008)$     & $(0.0002)$     \\
\hline
R$^2$       & $0.0093$       & $0.0190$       & $0.5589$       & $0.0474$       \\
Adj. R$^2$  & $-0.0379$      & $-0.0277$      & $0.5388$       & $0.0041$       \\
Num. obs.   & $23$           & $23$           & $24$           & $24$           \\
\hline
\multicolumn{5}{l}{\scriptsize{$^{***}p<0.001$; $^{**}p<0.01$; $^{*}p<0.05$}}
\end{tabular}
\caption{Statistical models for number of comments. See panels (a) and (b) of Figure~\ref{fig:toxicity_qr}.}
\label{table:reg_coefficients_qr1}
\end{center}
\end{table}

\begin{table}
\begin{center}
\begin{tabular}{l c c c c}
\hline
 & Model 1 & Model 2 & Model 3 & Model 4 \\
 & Questionable & Questionable & Reliable  & Reliable \\
 & (Real) & (Random) & (Real) & (Random) \\

\hline
(Intercept) & $0.1619$   & $0.3337^{***}$ & $0.2900^{***}$ & $0.3059^{***}$ \\
            & $(0.1552)$ & $(0.0822)$     & $(0.0142)$     & $(0.0198)$     \\
x           & $0.0165$   & $0.0023$       & $0.0078^{***}$ & $0.0018$       \\
            & $(0.0109)$ & $(0.0058)$     & $(0.0010)$     & $(0.0014)$     \\
\hline
R$^2$       & $0.0950$   & $0.0070$       & $0.7359$       & $0.0746$       \\
Adj. R$^2$  & $0.0539$   & $-0.0381$      & $0.7239$       & $0.0326$       \\
Num. obs.   & $24$       & $24$           & $24$           & $24$           \\
\hline
\multicolumn{5}{l}{\scriptsize{$^{***}p<0.001$; $^{**}p<0.01$; $^{*}p<0.05$}}
\end{tabular}
\caption{Statistical models for comments delay. See panels (c) and (d) of Figure~\ref{fig:toxicity_qr}.}
\label{table:reg_coefficients_qr2}
\end{center}
\end{table}

\end{document}